\documentclass[aps,showpacs,twocolumn,floats,prl]{revtex4}
\usepackage{graphicx}
\usepackage{amsmath}
\usepackage{psfrag}

\newcommand{\be}{\begin{equation}}              
\newcommand{\ee}[1]{\label{#1} \end{equation}}  

\newcommand{\bee}{\begin{eqnarray}}             
\newcommand{\eee}{\end{eqnarray}}               

\begin{document}

\title{Particle entrapment as a feedback effect}
\author{Sergey V. Shklyaev}
\affiliation{Theoretical Physics Department, Perm State
University, Bukirev 15, 614990 Perm, Russia}
\author{Arthur V. Straube}
\affiliation{Department of Physics, University of Potsdam, Am
Neuen Palais 10, PF 601553, D-14415 Potsdam, Germany}
\date{\today}
\begin{abstract}
We consider a suspension of polarizable particles under the action
of traveling wave dielectrophoresis (DEP) and focus on particle
induced effects. In a situation where the particles are driven by
the DEP force, but no external forces are exerted on the fluid,
the joint motion of the particles can induce a steady fluid flow,
which leads to particle entrapment. This feedback effect is proven
to be non-negligible even for small volume concentration of
particles.
\end{abstract}



\maketitle


%
%
Recent progress and numerous applications in medicine,
biotechnology and pharmaceutical research have witnessed a great
interest in understanding of fundamental aspects of particle
dynamics in fluid flows at small scales \cite{whitesides-06}.
Particularly, the concept of particle manipulation is an important
problem of micro- and nanofluidics \cite{squires-quake-05}. Here,
the particles can be physical (colloids, liquid droplets in
microemulsions, small bubbles) or biological (cells, bacteria,
biopolymers) objects manipulated with e.g., electrokinetic
\cite{morgan-green-03}, magnetic forces \cite{gijs-04}, ultrasound
\cite{hawkes-etal-04} or optical tweezers \cite{grier-03}. Despite
a considerable success in understanding of the impact of
hydrodynamic \cite{morgan-green-03, kim-karrila-91} and stochastic
\cite{brownian-motors} forces on the particle dynamics, the
problem of backward coupling, i.e., the influence of the particle
dynamics on the fluid, or {\it particle feedback}, includes a
number of open questions. Especially challenging is the problem of
{\it integral} feedback effects that can be induced by a
collection of {\it jointly} moving particles. In this Letter, we
systematically address this problem. We introduce an original
theoretical model, apply it to a physically realistic problem, and
predict a generic mechanism of particle entrapment that is
fundamentally different to the ones conventionally known
\cite{stommel-49, maxey-91, tuval-etal-05}.

We start with the formulation of a continuum model for
spherical noninteracting particles of radius $a$ suspended in a
fluid having viscosity $\eta$ and density $\rho$. To focus on the
feedback effects we impose a nonuniform external force
$F_0\mathbf{F}(\mathbf{r})$ on the particles ($F_0$ its reference
value and $\mathbf{F}$ is the dimensionless field), which,
however, does not directly influence the carrier phase. Provided
that the relative size of particles $a/L$ ($L$ is the length scale
of the flow) and their space-averaged volume fraction $\Phi_0$ are
small, the dynamics of the two-phase system is described by a
model allowing for the feedback \cite{model}:
%
%
%
\begin{eqnarray}
\frac{1}{\rm Sc}\left(\frac{\partial \mathbf{u}}{\partial
t}+\mathbf{u}\cdot \nabla \mathbf{u}\right) & = & -\nabla p+
\nabla^2\mathbf{u} +Q_{s}\left<\Phi\right>\varphi\mathbf{F}, \label{gov_eq1} \\
\rm{div}\,\mathbf{u} & = & 0, \quad
\mathbf{v}=\mathbf{u}+Q_{s}\mathbf{F}, \label{gov_eq2}\\
\frac{\partial \varphi}{\partial t}+\rm{div}\,\mathbf{j} & = & 0,
\quad \mathbf{j}=\varphi\mathbf{v}-\nabla\varphi, \label{gov_eq3}
\end{eqnarray}
where $\mathbf{u}$ and $\mathbf{v}$ are the fluid and particle
velocities, respectively, $p$ is the pressure, $\varphi$ is the
particle volume fraction, and $\mathbf{j}$ is the particle flux.
The equations have been nondimensionalized using the scales $L$,
$L^2/D$, $D/L$, $\eta D/L^2$, $\Phi_0$ ($D$ is the particle
diffusivity) for the length, time, velocity, pressure, and
particle volume fraction, respectively. Here $Q_s=2a^{2}L
F_0/9\eta D$ stands for the intensity of the external field, ${\rm
Sc}=\eta/D\rho$ is the Schmidt number, and a ratio of two
asymptotically small parameters
$\left<\Phi\right>=9L^2\Phi_0/2a^2$ is the feedback parameter,
which is assumed to be finite.

To stress the relevance of our approach hereafter we provide
estimations for a realistic system. As example, we stick to the
data close to experimental \cite{morgan-green-03}. For a water
($\eta \simeq 10^{-2} \;{\rm g}/{\rm cm\, s}$, $\rho \simeq 1
\;{\rm g/cm^3}$) suspension with $a \simeq 200 \;{\rm nm}$ and
$\Phi_0 \simeq 3\%$ in a container of the size $2L \simeq 25 \;
\mu{\rm m}$ at temperature $300\,{\rm K}$ one obtains
$\left<\Phi\right> \simeq 500$ and according to Einstein's formula
$D \simeq 10^{-8} \; {\rm cm}^2/{\rm s}$ and ${\rm Sc} \simeq
10^6$. Although in most conventional situations ${\rm Sc}$ is
high, it becomes necessary to account for the diffusion of
particles. The reason is twofold: (i) even small diffusion gets
non-negligible at small scales, e.g., the diffusion time is of
order $L^2/D \simeq 100 \;{\rm s}$; (ii) it prevents from
unbounded physically irrelevant accumulation of particles by the
external field.

We emphasize that only particles are able to make fluid move,
i.e., the fluid flow itself is a perfect indicator for the
particle feedback, described by the last term in
Eq.~(\ref{gov_eq1}). Physically, this term comes from the Stokes
drag, which dominates in the interphase force \cite{maxey-91,
druzhinin-95} and is balanced by $\mathbf{F}$. Particularly, this
results in a distinction of the velocities of phases, where
inertia corrections are negligible.

Further we focus on an example of a dielectophoretic (DEP) force
exerted on polarizable particles under ac electric field
$\mathbf{E}(\mathbf{r},t)={\rm
Re}[\tilde{\mathbf{E}}(\mathbf{r})\exp(i\omega t)]$. Hereafter
$\omega$ is the angular frequency, $i=\sqrt{-1}$, ${\rm
Re}[z]\equiv z_r$ and ${\rm Im}[z]\equiv z_i$ denote the real and
imaginary parts of $z$. The time-averaged force (per unit volume)
reads \cite{morgan-green-03}:
\begin{equation}\label{DEP_force}
F_0\mathbf{F}=\frac{3}{2} \epsilon_m {\rm Re} \left[\tilde
K(\omega)\tilde{\mathbf{E}} \cdot \nabla
\tilde{\mathbf{E}}^{*}\right],
\end{equation}
where ``$^{*}$'' indicates complex conjugate. The complex
frequency-dependent function $\tilde K(\omega)=(\tilde
\epsilon_p-\tilde \epsilon_m)/(\tilde \epsilon_p+2\tilde
\epsilon_m)$ is a measure of an effective polarizability of the
particle, known as the Clausius-Mossotti factor. Here,
$\tilde{\epsilon}_p$ and $\tilde{\epsilon}_m$ are the complex
permittivities of the particles and the fluid medium,
respectively. The complex permittivity is defined as $\tilde
\epsilon=\epsilon-i\sigma/\omega$ ($\epsilon$ is the permittivity
and $\sigma$ is the conductivity of the dielectric). The DEP force
(\ref{DEP_force}) comprises two independent contributions:
%
\begin{equation}\label{}
F_0\mathbf{F}=\frac{3}{4} \epsilon_m \tilde{K}_r \,\nabla
|\tilde{\mathbf{E}}|^2-\frac{3}{2} \epsilon_m\tilde{K}_i\,\nabla
\times \left(\tilde{\mathbf{E}}_r\times
\tilde{\mathbf{E}}_i\right).
\end{equation}
The first term relates to the in-phase component of the induced
dipole. This force points towards the domains of higher field
strength for $\tilde{K}_r>0$ or, conversely, to the domains of
weaker fields for $\tilde{K}_r<0$, which is referred to as {\it
positive}-DEP ({\it p}-DEP) or {\it negative}-DEP ({\it n}-DEP),
respectively \cite{pohl-78}. The particles are attracted or
repelled by the electrode edges. The second term is due to the
out-of-phase component of the dipole and is essential if there is
spatially varying phase \cite{morgan-green-03}, e.g., for
traveling wave DEP; it makes the particles move parallel to the
electrodes.

To be able to focus on the feedback effects one has to carefully
ensure that these are not hindered by other possible sources of
fluid motion. Because of Joule heating, applied electric fields
induce temperature gradients and therefore create nonuniformities
of the conductivity, permittivity and density in the fluid, which
can lead to electroconvection and/or natural convection
\cite{morgan-green-03}. A typical electrothermal force on the
fluid $F_{ET}\propto \sigma_m U_0^4 L^{-3}$ is negligible compared
to the feedback term, provided that $F_{ET}/F \propto \sigma_m
U_0^2 /\Phi_0 \ll 1$ ($U_0$ is a characteristic value of the
electric potential). This requirement can be safely satisfied for
weak fields or weakly conductive fluids. Further we assume that
the last option is met, e.g., for pure water ($\sigma_m \simeq 30
\;{\rm \mu S/cm}$), or deionized water ($\sigma_m \simeq 2 \;{\rm
\mu S/cm}$) applied in \cite{tuval-etal-05}, and $U_0 \simeq 0.1
\;{\rm V}$ we obtain $F_{ET}/F$ of order $10^{-5}$ or $10^{-6}$,
respectively. As we claim below, considerably weaker fields are
enough to cause the feedback-induced flow (cf. with $U_0 \simeq
1-10\;{\rm V}$ in \cite{morgan-green-03,maika,tuval-etal-05}). One
more advantage of the low conductivity is that ac electro-osmosis
(another electrohydrodynamic effect that can induce a flow caused
by electrical stresses in the diffuse double layer of charges near
the electrodes) and natural convection are even weaker than
electroconvection.

We now turn to the analysis of a system typical for experiments on
traveling wave DEP \cite{morgan-green-03, maika}. Consider the
two-phase medium filling a rectangular container of sizes $L_x$,
$L_y \equiv 2L$, $L_z$ and impose a traveling wave of the
potential at the boundaries $y=\pm L$: $\phi=U_0 \exp[i(\omega
t-qx)]$, where $q$ is the wave number and $U_0$ is the amplitude.
The complex amplitude $\tilde\phi(\mathbf{r})$
($\tilde{\mathbf{E}}=-\nabla\tilde\phi$) obeys the Laplace
equation, $\nabla^2\tilde\phi=0$, which is readily solved.
Assuming that $L_y \ll L_x$, $L_y \ll L_z$ and that the sidewalls
are electrically passive, we obtain
$\tilde\phi(\mathbf{r})=U_0\exp(-iqx) (\cosh qy)/\cosh qL$ and
evaluate
\begin{equation}\label{force}
\mathbf{F}(\mathbf{r})=\left(-K_i \cosh by, \, \sinh by, \,
0\right).
\end{equation}
Here we define $F_0=3\epsilon_m U_0^2 q^3 \tilde K_r /2 \cosh^2
(b/2)$ and introduce a dimensionless parameter $K_i=\tilde
K_i/\tilde K_r$ and the dimensionless wavenumber $b=2qL$. A
traveling wave of a period $50 \; \mu{\rm m}$ leads to $b\approx
3$ and $|Q_s| \approx 0.5$ (estimations are based on the data as
before). Next, we restrict our consideration to the case of
$K_i>0$ as reversal of the sign of $K_i$ changes the direction of
the induced flow.

We first point out to a partial case
of $K_i=0$, which corresponds to the limit of perfectly lossless
dielectric particles. According to (\ref{force}), there is no
force allowing for the particle transport along the plates $y=\pm
1$, only transversal redistribution occurs. The particles tend to
migrate either towards or away from these boundaries, which is
counterbalanced by diffusion; longitudinal nonuniformities are
smeared by diffusion. Thus, this situation admits a state of {\it
mechanical equilibrium} described by the quiescent fluid
$\mathbf{u}_0(\mathbf{r})=0$, vanishing particle flux
$\mathbf{j}_0(\mathbf{r})=0$, and a nonuniform distribution of
particles
\begin{equation}\label{ME-distrib}
\varphi_0(\mathbf{r})=C_0\Psi_0(y), \quad
\Psi_0(y)=\exp\left(Q_s b^{-1} \cosh by \right),
\end{equation}
\noindent where $C_0=1/\int_0^{1}\Psi_0(y)dy$, as by definition
the averaged $\varphi$ is unity. The concentration profile
(\ref{ME-distrib}) describes accumulation of particles near the
boundaries for $Q_s>0$, or in the center plane for $Q_s<0$, which
corresponds to {\it p}-DEP or {\it n}-DEP, respectively.
Analytical and numerical treatment of the linearized problem as
well as a direct numerical simulation (DNS) of the nonlinear model
(\ref{gov_eq1})-(\ref{gov_eq3}) with (\ref{force}) indicate that
the state of mechanical equilibrium is stable for any values of
the governing parameters.

\begin{figure}[!t]
\centering \includegraphics[width=0.45\textwidth]{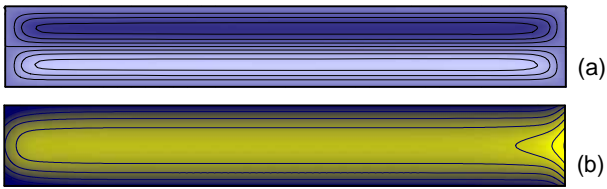}
\caption{(Color online). Contourplots of the streamfunction (a)
and particle concentration (b) in the steady state for ${\rm
Sc=2}$, $Q_s=-0.2$, $K_i=1$, $b=3$, $\left<\Phi\right>=575$.
Lighter (clockwise rotation for vortices) and darker
(counterclockwise rotation) colored domains refer to higher and
lower values of the plotted fields, respectively. Steady states
corresponding to different ${\rm Sc}$ look similar: a slight
distinction comes from the nonlinear term, which is nonvanishing
only near the sidewalls. } \label{fig1:closed-box}
\end{figure}

What happens in a more
general case of lossy particles, when $K_i \ne 0$ and the
longitudinal transport is allowed, is a simple question to pose,
but the remarkably hard one to answer. In the limiting case of no
feedback, $\left<\Phi\right> \ll 1$, there is no source for fluid
motion and the problem reduces to finding a distribution
$\varphi_0(\mathbf{r})=\varphi_0(x,y)$, governed by
Eqs.~(\ref{gov_eq2}), (\ref{gov_eq3}) with
$\mathbf{u}(\mathbf{r})=0$. In the presence of the feedback, the
problem becomes highly nontrivial, because the system is
mechanically nonequilibrium. To get an impression of possible
scenarios, we have numerically integrated
Eqs.~(\ref{gov_eq1})-(\ref{gov_eq3}) with (\ref{force}) in a
two-dimensional rectangular box with the no-slip condition for
$\mathbf{u}$ and vanishing normal component for $\mathbf{j}$ at
the solid walls. A typical steady state solution is presented in
Fig.~\ref{fig1:closed-box}. The flow is of a large scale and
closed, the particles are involved in the vortical motion, which
is reminiscent of the particle entrapment under gravity by Stommel
\cite{stommel-49} (for high aspect ratio see Ref.
\onlinecite{capture-05}). However, there are two principal
differences. First and most important, the conventional entrapment
implies existence of a vortex flow irrespective of whether there
are any particles or not \cite{stommel-49, maxey-91,
tuval-etal-05, capture-05}. The fluid flow in our system can be
induced only by means of particles and is not possible otherwise.
In contrast to the previous studies, particle entrapment arises as
a generic particle feedback effect, which also provides a way to
generate a flow. The second important distinction is that we
carefully account for the diffusion effects. This nontrivial
problem has been systematically addressed neither in studies cast
into a Hamiltonian frame \cite{stommel-49, maxey-91, capture-05},
nor in a non-Hamiltionan system \cite{tuval-etal-05}.
%
%
%
%

A closer inspection of Fig.~\ref{fig1:closed-box} shows that the
flow and concentration patterns are one dimensional (1D)
everywhere except for the vicinities of the sidewalls. This allows
for a 1D analysis of Eqs.~(\ref{gov_eq1})-(\ref{gov_eq3}) with
(\ref{force}) away from the sidewalls, valid for systems with a
high aspect ratio. Accordingly, we apply the ansatz
$\varphi_0({\mathbf r})=\varphi_0(y)$, ${\mathbf u}_0({\mathbf
r})=(u_0(y),0,0)$. As the flow does not influence the distribution
of particles, $\varphi_0(y)$ is given by
(\ref{ME-distrib})
with the same $C_0$. To determine the fluid velocity we account
for the no-slip conditions at the solid walls $u_0(\pm 1)=0$ and a
condition of no mean fluid flux $\int_{-1}^{1}u_0 dy =0$, implying
that the flow is closed. As a result, we obtain:
\begin{equation}
u_0(y)=\alpha V_0(y), \quad V_0(y)=V_{01}(y)+V_{02}(y)
\label{MNE-u0}
\end{equation}
with $V_{01}(y)=I_2(1)-I_2(y)$, $V_{02}(y)=-\beta(1-y^2)$, where
$\alpha=C_0 \left<\Phi\right> |Q_s K_i|>0$,
$\beta=3[I_2(1)-I_3(1)]/2>0$, $I_1(y)=\int_0^y\Psi_0(\xi)\cosh
b\xi d\xi$, and $I_{l+1}(y)=\int_0^y I_l(\xi) d\xi$ $(l=1,2)$;
because of symmetry, $V_0(y)$ is an even function. Next, we
impose the condition of particle entrapment
$\int_{-1}^{1}\mathbf{j}_0 \cdot \mathbf{e}_x \,dy=0$,
$\mathbf{e}_x=(1,0,0)$, which ensures no mean particle flux
\cite{capture-05}. We arrive at
%
\begin{equation}\label{Phi_lim}
\left<\Phi\right>=I_1(1)\left( C_0 \int_0^1 V_0 \Psi_0
dy\right)^{-1}\equiv \left<\Phi\right>_c,
\end{equation}
representing a formal restriction on $\left<\Phi\right>$: for
every set of governing parameters, only a specific number of
particles defined by (\ref{Phi_lim}) can be maintained trapped by
the flow. In practice, however, e.g., for a closed rectangular
box, such a restriction is not stringent. For $\left<\Phi\right>$
different from $\left<\Phi\right>_c$, the solution away from the
sidewalls still corresponds to (\ref{ME-distrib}), as if
$\left<\Phi\right>=\left<\Phi\right>_c$. Their actual distinction
is balanced in the vicinities of the sidewalls, where the lack or
excess of particles emerges leading to local gradients of
concentration on top of (\ref{ME-distrib}). This is clearly seen
in Fig.~\ref{fig1:closed-box}(b). Of special attention is the case
of $\left<\Phi\right>$ considerably smaller than
$\left<\Phi\right>_c$. Here, although there are not enough
particles to excite the flow in the whole domain, the entrapment
still occurs. As before, it is accompanied by the birth of a
steady vortex flow, but of a smaller longitudinal extension. With
the decrease of $\left<\Phi\right>$ the vortices gradually shrink
and in the limit of $\left<\Phi\right> \ll 1$ no longer exist.

%
%
\begin{figure}[!t]
\centering $\;$\includegraphics[width=0.45\textwidth]{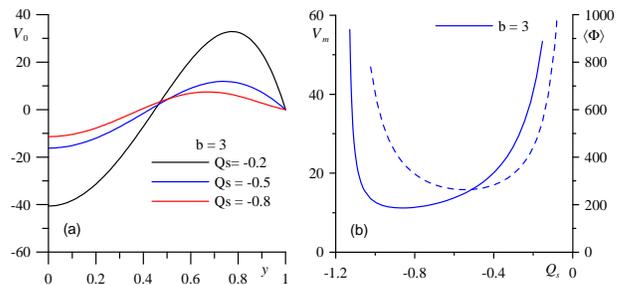}
\caption{(Color online). Characteristics of 1D state of entrapment
(\ref{ME-distrib})-(\ref{Phi_lim}) for $b=3$: velocity profiles
(a), maximal absolute velocity $V_m=\max_{y}|V_0(y)|$ and
$\left<\Phi\right>_c$ as functions of $Q_s$ (b).}
\label{fig2:base-state}
\end{figure}
%

The characteristics of the discussed state are presented in
Fig.~\ref{fig2:base-state}. The particles tend to move along the
axis $x$, faster near the boundaries and slower at the center of
the channel [see Eqs. (\ref{gov_eq2}) and (\ref{force})]. Because
of viscous drag, the fluid is towed by the particles, which has
two consequences. First, this motion contributes in a positive
fluid flux, defined by $V_{01}(y)$. Second, it creates a
longitudinal gradient of pressure that gives rise to an opposite
Poiseuille flow $V_{02}(y)$ with the maximal velocity at the
center. The velocity profile $V_{0}(y)$ is a superposition of
these counterflows, such that the net fluid flux is vanishing.
Noteworthy, profile (\ref{MNE-u0}) qualitatively resembles the one
in the convective flow in a vertical slot induced by internal
sources of heat \cite{gershuni-zhukhovitsky-76}. In our case, the
role of the heat sources is played by the nonuniform distribution
of particles and the DEP force instead of gravity. Note, the state
of entrapment exists only for $Q_c<Q_s<0$, see
Fig.~\ref{fig2:base-state}(b). Beyond this range,
Eq.~(\ref{Phi_lim}) prescribes negative values of
$\left<\Phi\right>_c$, which is physically irrelevant.
Particularly, this manifests that the entrapment can be observed
only for {\it n}-DEP. From the experimental point of view, the
flow can be easily controlled by tuning the frequency of the
imposed traveling wave.

\begin{figure}[!t]
\centering \includegraphics[width=0.48\textwidth]{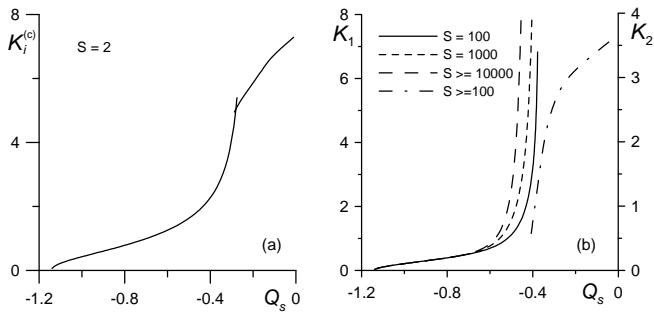}
\caption{Stability map for $b=3$, ${\rm Sc=2}$ (a). Scaling
functions $K_1(Q_s)$ and $K_2(Q_s)$ for different ${\rm Sc}$ (b).}
\label{fig3:stability}
\end{figure}

Next question concerns existence of the revealed effect in real
systems. To answer it, we have linearized
Eqs.~(\ref{gov_eq1})-(\ref{gov_eq3}) near solution
(\ref{ME-distrib}), (\ref{MNE-u0}) and investigated its stability
with respect to perturbations of the form
$f(\mathbf{r})=\hat{f}(y)\exp(\lambda t - i k_x x-i k_z z)$. Here
$\lambda=\lambda_r+i\lambda_i$ is the complex growth rate, $k_x$
and $k_z$ are the wave numbers along axes $x$ and $z$,
respectively. The analysis has shown that the modes with the
largest $\lambda_r$ correspond to the perturbations in the form of
rolls, $k_z=0$. We have checked a wide range of Schmidt numbers
$1<{\rm Sc}<10^{6}$ and found no qualitative changes, the
stability maps are presented in Fig.~\ref{fig3:stability}. The
regions of stable and unstable behavior are separated by two
curves $K_i^{(c)}(Q_s)$ of neutral stability, on which
$\lambda_r=0$. These lines refer to a pair of competing modes of
the largest $\lambda_r$ and have different asymptotes for ${\rm
Sc}\gg 1$. For the branches with higher and lower $|Q_s|$ the
scaling laws are $K_i^{(c)}=K_1 \sqrt{\rm Sc}$ and $K_i^{(c)}=K_2
\,{\rm Sc}$, respectively. The dependencies $K_{1}(Q_s)$ and
$K_{2}(Q_s)$ for different ${\rm Sc}$ are plotted in
Fig.~\ref{fig3:stability}(b). With the growth of ${\rm Sc}$, these
dependencies converge to master curves. Because of the different
scaling, the convergence of $K_{1}$ is slower, whereas $K_2$ gets
indistinguishably close to its master curve already at ${\rm
Sc}=100$.
%
%
%
%

\begin{figure}[!h]
\centering \includegraphics[width=0.45\textwidth]{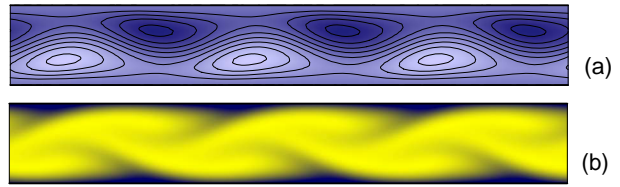}
\caption{(Color online). Breakdown of 1D state beyond the
stability threshold, ${\rm Sc=2}$, $Qs=-0.2$, $K_i=7$, $b=3$,
$\left<\Phi\right>=450$: streamfunction (a) and particle
concentration (b).} \label{fig4:broken_state}
\end{figure}

The results of the linear stability analysis were confirmed by DNS
of the nonlinear model (\ref{gov_eq1})-(\ref{gov_eq3}) with
(\ref{force}). Note, as for large ${\rm Sc}$ the value $K_i^{(c)}$
is high, the instability in this particular situation is hardly
reachable experimentally. However, it can be the case for moderate
${\rm Sc}$. We have studied the breakdown of the 1D state, which
is found to occur supercritically. The patterns beyond the
threshold (see Fig.~\ref{fig4:broken_state} for a snapshot) travel
along the axis $x$ with a speed $\lambda_i/k_x$. For the different
branches the patterns look similar, but have distinct spatial
periods and travel in opposite directions.

In conclusion, we have studied the role of the particle feedback
in a two-phase system under the action of traveling wave DEP. In a
situation where the particles are driven by the DEP force but no
external forces are exerted on the fluid, the joint motion of the
particles can induce a steady fluid flow, which is accompanied by
novel particle entrapment. In a contrast to the conventional
mechanism, diffusion of the particles becomes a necessary
ingredient for the entrapment. This particle feedback effect has
been proven to be non-negligible even for small volume
concentration of particles.
%
We note that similar phenomena are expected to exist in various
physical systems. Indeed, the set of
Eqs.~(\ref{gov_eq1})-(\ref{gov_eq3}) with the force in the form
(\ref{DEP_force}) describe a wide class of problems, e.g.,
magnetized ferrofluids \cite{shliomis-smorodin-02}, particles
driven by optical tweezers \cite{grier-03}, bubbly fluids under
vibrations \cite{vibro-06}, where the field $\tilde{\mathbf{E}}$
entering (\ref{DEP_force}) is of different nature.
%
%
%
%

We acknowledge fruitful discussions with A.A.~Nepomnyashchy,
A.~Pikovsky, B.~L.~Smorodin, V.~Steinberg, and C.~Pooley. S.S.
thanks DAAD for support; A.S. was supported by the German Science
Foundation (DFG, SPP 1164 ``Nano- and microfluidics,'' project
1021/1-1).



\begin{thebibliography}{10}
\bibitem{model}
Details will be published elsewhere; for approach see Refs.~\onlinecite{druzhinin-95,capture-05, vibro-06}
\end{thebibliography}
\end{document}